\documentclass[floatfix,twocolumn,prl,showpacs,aps,superscriptaddress]{revtex4}
\usepackage{graphicx}
\usepackage{subfigure}
\usepackage{amssymb}
\usepackage{amsmath}
\usepackage[dvips]{epsfig}
\usepackage[usenames]{color}

\def\del{\nabla}
\def\n{\textbf{n}}
\def\N{\textbf{N}}
\def\k{\bar{\kappa}}
\def\qsm{q_{\textrm{sm}}}

\def\sn{\textrm{sn}}
\def\cn{\textrm{cn}}
\def\dn{\textrm{dn}}

\def\thex{\tilde{\Theta}_{\textrm{hex}}}
\def\TGB{\textrm{TGB}}

\begin{document}

\title{Helical Nanofilaments and the High Chirality Limit of Smectics-A}

\author{Elisabetta A. Matsumoto}\affiliation{Department of Physics and Astronomy, University of Pennsylvania, 209 South 33rd Street, Philadelphia, Pennsylvania 19104-6396, U.S.A.}
\author{Gareth P. Alexander}\affiliation{Department of Physics and Astronomy, University of Pennsylvania, 209 South 33rd Street, Philadelphia, Pennsylvania 19104-6396, U.S.A.}
\author{Randall D. Kamien}
\affiliation{Department of Physics and Astronomy, University of Pennsylvania, 209 South 33rd Street, Philadelphia, Pennsylvania 19104-6396, U.S.A.}\affiliation{School of Mathematics, Institute for Advanced Study, Princeton, NJ 08540, U.S.A.}

\date{\today}
\pacs{61.30.-v, 61.30.Mp, 64.70.M-}
 
\begin{abstract}
Liquid crystalline systems exhibiting both macroscopic chirality and smectic order experience frustration resulting in mesophases possessing complex three-dimensional order. In the twist-grain-boundary phase, defect lattices mediate the propagation of twist throughout the system. We propose a new chiral smectic structure composed of a lattice of chiral bundles as a model of the helical nanofilament (B4) phase of bent core smectics.
\end{abstract}

\maketitle

The chiral nature of bulk liquid crystalline textures is accompanied by a form of inherent frustration as the local preferred configuration is often incompatible with global, or topological, requirements~\cite{kamien01b}. In such cases, the chirality leads to spectacular mesophases, for example the cholesteric blue phases~\cite{wright89} or the twist grain boundary (TGB) phase~\cite{renn88}, involving the proliferation of topological defects to relieve the frustration. The expression of macroscopic chirality does not always result from molecular chirality, but can also arise from a spontaneously broken symmetry in achiral bent core liquid crystals~\cite{brand05,link97,sekine97,yan08}. A number of novel, intricate textures form in these bent core systems, such as the opposite handed domains of the dark conglomerate phase and, of particular focus here, the recently discovered helical nanofilament (HN) phase, also known as the B4 phase~\cite{hough09a}.
The hierarchical structure of the HN phase is built upon helical bundles, each comprised of roughly five layers, which independently nucleate, all with the same handedness, and coalesce to form a bulk texture of parallel, coherently rotating nanofilaments, resulting in homochiral domains that appear on the scale of tens of microns. Remarkably, the texture of the HN phase does not correspond to the traditional TGB phase, suggesting the existence of a new class of solutions in the venerable problem of chiral smectics~\cite{noel}. In this Letter we propose a new class of solutions, consistent with the observed structures and appropriate when the intrinsic chirality is high. 

We begin by re-examining the Landau-de Gennes theory as a means of describing the macroscopic chirality. The free energy is given by the minimal coupling of the complex smectic order parameter $\psi=\psi_0 \text{e}^{i q_{\textrm{sm}} \phi}$ to the liquid crystal director ${\bf n}$~\cite{degennes72} 
\begin{eqnarray}
F &=& \int d^3x \biggl\{ C \vert (\nabla - iq_{\text{sm}}{\bf n})\psi \vert^2 + (t-t_c) \vert \psi \vert^2 + \frac{u}{4} \vert\psi \vert^4\nonumber\\
&&\qquad +\frac{K_1}{2}(\del \cdot \n)^2+ \frac{K_2}{2} \bigl[ \n \cdot (\del \times \n) + q_0 \bigr]^2\nonumber\\
&&\qquad + \frac{K_3}{2} \bigl[ (\n\cdot\del)\n \bigr]^2 + f_{\textrm{G}} \biggr\}
\label{eq:ldg_energy}
\end{eqnarray}
where $a=2 \pi/q_{\textrm{sm}}$ is the layer spacing, $q_0$ is the chirality, the $K_i$ are Frank elastic constants and the bending energy of the layers due to Gaussian curvature $K$ is $f_{\textrm{G}}= \bar{\kappa} \nabla\cdot\left[ (\N \cdot \del)\N - \N(\del \cdot \N) \right]$, where $\k$ is the bending modulus and $\N$ is the layer normal.
The coupling term decomposes into gradients in the magnitude of the order, which are neglected in the London limit, and the layer compression energy, $\tfrac{B}{2} \vert \del \phi - \n \vert^2$, where $B=2 C \qsm^2 \psi_0^2$ is the compression modulus. It is important to note that the Gaussian curvature energy is typically neglected as it is a boundary term, however, additional boundaries caused by defects in the system make this term relevant.

\begin{figure}[t]
\begin{center}
\includegraphics[width=80mm]{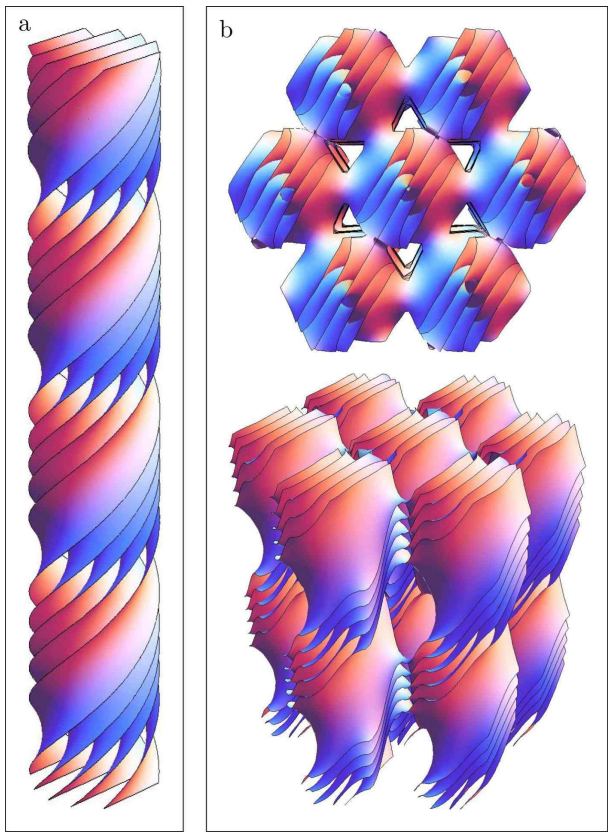}
\caption{(Color online) a) A single bundle of helicoidal layers locally attains the absolute minimum of the chiral Landau-de Gennes free energy. b) As many bundles nucleate, they assemble into a lattice with defects.}
\label{smecticbundles}
\end{center}
\end{figure}

Much of the phenomenology of chiral smectics follows from de Gennes' observation that the gauge-like nature of the coupling term leads to a formal analogy between smectics and superconductors~\cite{degennes72} and, indeed, the equilibrium phases are in close correspondence. The nature of the equilibrium textures is controlled by the ratio of the penetration depth $\lambda=(K_2/B)^{1/2}$ to the correlation length $\xi= (C/\vert t_c-t \vert)^{1/2}$. In a type I material, $\tfrac{\lambda}{\xi} < \tfrac{1}{\sqrt{2}}$, the only low temperature phase is the smectic-A in which the layers are flat and the chirality is fully expelled up to a thermodynamic critical value $q_{\text{th}} = (\tfrac{2}{K_2u})^{1/2} |t_c-t|$, at which there is a first order transition to the cholesteric $N^*$ phase. In type II materials, $\tfrac{\lambda}{\xi} > \tfrac{1}{\sqrt{2}}$, it becomes possible for the twist to penetrate at least partially in defect proliferated states analogous to the Abrikosov flux vortex phase in superconductors~\cite{abrikosov57}. 

The twist grain boundary phase represents one solution of this kind in which locally flat layers rotate through a fixed angle $\alpha$ across the boundary defined by an infinite line of screw dislocations~\cite{renn88,kamien99}. For highly chiral systems we show that a second solution exists, that is based upon an underlying cholesteric director field and motivated in part by both the coherent rotation and the Bouligand texture of the filaments observed experimentally \cite{hough09a}.

We seek a high chirality solution 
in which the director retains its high temperature cholesteric form, ${\bf n} = \cos (q_0z){\bf e}_x + \sin (q_0z){\bf e}_y$, and the smectic phase field is then chosen to minimize the compression energy for this choice of director. Since $\nabla\times{\bf n} = -q_0{\bf n}$, there is no phase field for which $\nabla \phi = {\bf n}$ everywhere, however, we can construct a phase field that agrees on a lower dimensional set of points. To this end, we consider the local phase field 
\begin{equation}
\phi^{\text{loc}} = x \cos (q_0z) + y \sin (q_0z) = r \cos (q_0z-\theta) ,
\label{eq:phi_local}
\end{equation} 
where in the final equality we have employed cylindrical coordinates. The compression is $\vert \nabla \phi^{\text{loc}} - {\bf n} \vert^2 = q_0^2r^2\sin^2(q_0z-\theta)$ and vanishes everywhere on the two-dimensional surface of the helicoid $\theta = \{ q_0z,q_0z+\pi \}$. This texture, therefore, locally attains the absolute minimum of the Landau-de Gennes free energy and represents the optimum local configuration for a chiral smectic. The smectic layers (Fig.~\ref{smecticbundles}.a) are themselves helicoids, the level sets of $\phi^{\text{loc}}$, that intersect the zero compression surface orthogonally.
Consequently the compression of the layers grows quadratically with their radial size so that they are naturally confined to a finite region; the texture is only local and does not fill space. The free energy per unit length of a cylindrical region of radius $R$ is $\pi Bq_0^2R^4/8\,$. A comparison with the free energy per unit length of an equal volume region of the smectic-A texture, $\pi K_2q_0^2R^2/2$, suggests a natural size of $R_* = \sqrt{2} \lambda$. 

This type of local construction is reminiscent of double twist cylinders in the blue phases~\cite{wright89}: there too the absolute minimum of the free energy density is attained only locally, on the axes of the double twist cylinders. Space filling textures, the cubic blue phases, are constructed by repeating this local texture in a periodic manner at the expense of incorporating topologically required disclination lines. We seek an analogous construction for a bulk texture of our helical smectic bundles. Just as in the blue phases, such a state will be stable if the benefit gained from the locally favored bundles outweighs the cost of any associated defects. 

It is natural to consider two-dimensional arrangements of bundles since then they can all be supported by a common underlying cholesteric director field -- such a construction automatically incorporates the macroscopic phase coherence evidenced by the Bouligand texture of the experiment~\cite{hough09a}. 
The smectic phase field can be obtained by constraining its form to match the rotation of the director along the $z$-direction and considering only variations with respect to $x,y$. The constrained Euler-Lagrange equation is then $\nabla_{\!\perp}\! \cdot (\psi_0^2\nabla\phi)=\nabla_{\!\perp}\!\cdot(\psi_0^2{\bf n})$, so that in the London approximation where $\psi_0$ is constant almost everywhere $\phi$ is harmonic in $x,y$; we propose the general {\it ansatz} for the phase field $\phi = \text{Re}[\Theta (w) \text{e}^{-iq_0z}]$, where $\Theta$ is an analytic function of $w=x+iy$.

The local optimum configuration, Eq.~\eqref{eq:phi_local}, is recovered when $\Theta \sim w$ is linear. A bulk texture will therefore correspond to an analytic function with a large number of simple zeros in the complex plane, as each defines the axis of a helical bundle. Lattice configurations of simple zeros can be easily constructed with the aid of elliptic functions. It is convenient to introduce a dimensionless coordinate $\zeta = w/l$, where $l$ is the lattice constant. Our analysis here focuses solely on a triangular lattice of bundles (Fig.~\ref{smecticbundles}b), given by 
\begin{equation}
\tilde{\Theta}_{\textrm{hex}}(\zeta)=\frac{3^{1/4}}k \left(1+\sqrt{3} \frac{\cn^2(k\zeta,m)}{\dn^2(k\zeta,m)\sn^2(k\zeta,m)}\right)^{-1/2},
\label{eq:hex2}
\end{equation}
where $k = 2K(m) \approx 3.196$ is twice the complete elliptic integral of the first kind and $m=\frac{2-\sqrt{3}}4$ is the square of the elliptic modulus that sets the period ratio. The zeros of $\tilde{\Theta}_{\textrm{hex}}$ sit on a triangular lattice, and the complementary divergences take the form of square root branch points, $\tilde{\Theta}_{\textrm{hex}} \sim \zeta^{-1/2},$ situated on a honeycomb lattice. Simple zeros correspond to helicoids of the same handedness and pitch as the cholesteric director field while the layer structure near the divergences is that of a half-helicoid of the opposite handedness. Adjacent bundles join together smoothly through saddle points, at which the bulk structure deviates significantly from the single bundle structure.

At these saddle points, we might expect the director field to deviate from the simple cholesteric form so as to more closely follow the layer normal, reducing the compression energy.
Such screening effects are crucial near the lower critical chirality of the $\TGB$~\cite{renn88} and should be included when the characteristic length scale of the structure is large compared to the penetration depth. However, both the compression and twist contributions are relevant to the free energy, suggesting that we expect $K_2 q_0^2 \sim  B$ or equivalently $q_0 \lambda \gtrsim 1$. In the experiment $q_0l \sim 1$~\cite{hough09a}, suggesting that $l \lesssim \lambda$ so that neglecting screening effects should be appropriate. 
\begin{figure}[t]
\begin{center}
\includegraphics[width=45mm]{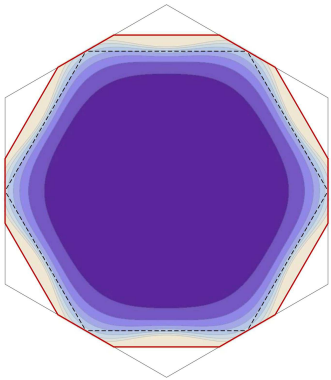}
\caption{(Color online) In the contour plot of the compression energy density for a single hexagonal unit cell, we approximate the region $\tilde{\Omega}$ where smectic order exists by the truncated hexagon bounded by the solid red line.  Within the dashed black line the compression energy is well approximated by an expansion about the origin, while outside it an expansion about the singularities is used.}
\label{energydensity}
\end{center}
\end{figure}
Working in the London limit, the free energy per unit volume of the bulk texture becomes 
\begin{eqnarray}
\frac{F}{V} &=& \frac{2}{\sqrt{3}} \int_{\tilde{\Omega}} d^2\zeta  \Bigl\{ \frac{Bq_0^2l^2}{4} |\tilde{\Theta}_{\textrm{hex}}|^{2}\nonumber\\
&&+\frac{B}{2} \bigl| \partial_\zeta \tilde{\Theta}_{\textrm{hex}}- 1 \bigr|^2 -\frac{(t-t_c)^2}{u}\Bigr\},
\label{eq:hexenergy}
\end{eqnarray}
where we have integrated over the pitch direction, $(t-t_c)^2/u$ is the condensation energy,  and $\tilde{\Omega}$ is the non-dimensionalized domain (Fig. \ref{smecticbundles}.c) in which smectic order is present. The compression energy diverges at singularities in $\thex$, forcing the smectic order to vanish, giving the bulk texture a porous appearance. The energy within the central region defined by the hexagon joining the midpoints between divergences is dominated by the Taylor expansion of $\tilde{\Theta}_{\textrm{hex}} = \zeta - \tfrac{|k|^6}{42\sqrt{3}} \zeta^7 + O(\zeta^{13})$ about the origin. Similarly, the energy near the defects is dominated by the Taylor series about the singularities $\tilde{\Theta}_{\textrm{hex}} = c(1-i)\zeta^{-1/2} + O(\zeta^{5/2})$, where $c=\frac{1}{2} 3^{3/8}k^{-3/2} \approx 0.132$, leading to an energy per unit volume of 
\begin{eqnarray}
\frac{F}{V} & =& B \biggl[q_0^2 l^2 \Bigl( \frac{5}{256}+ c^2 \cosh^{-1}(2)\bigl( 1 - 2(1 - \epsilon)^{1/2} \bigr) \Bigr)\nonumber \\ 
&+&0.052 + 6 \sqrt{3}c^2 \bigl( (1 - \epsilon)^{-1/2} - 2 \bigr) \biggr] - \epsilon \frac{(t-t_c)^2}{u}, \quad
\label{eq:hexenergy_numbers}
\end{eqnarray} 
where $\epsilon$ is the filling fraction and the result is valid in the regime where adjacent bundles overlap, $\epsilon \in [\tfrac{3}{4},1)$~\cite{filling}. For the texture to be stable the divergence in the compression must not be more costly than the condensation energy. Assuming the coherence length is comparable to the layer spacing, $q_{\textrm{sm}} \xi \sim 1,$ then sets a hard upper bound for the filling fraction of $\epsilon < 0.91$. It follows that the regions of suppressed smectic order are substantial and occupy a finite fraction of the unit cell.

We note that the only dependence on the lattice constant in the free energy density, Eq.~\eqref{eq:hexenergy}, is quadratic in $l$. Since the coefficient is positive definite, the energy expression would lead one to conclude that $l$ should shrink to zero~\cite{Sawtooth}. 
Of course the small-scale features will be cut off by microscopic length scales as well as additional energetic considerations. Recalling that bent-core mesogens favor layers with negative Gaussian curvature~\cite{hough09a}, we consider the saddle-splay of the individual layers as the bundle size shrinks: since the saddle-splay is the same as $K$~\cite{didonna02}, when the Frank constant $\k<0$ is negative, the system rewards the negative $K$ everywhere in each layer. The Gauss-Bonnet theorem states that the integrated Gaussian curvature is purely a topological quantity. Thus, each layer in a bundle contributes the same free energy, $\frac{F_G}V \sim \frac{9}{16} \k q_0^2$ for a filling fraction of $\epsilon =3/4$. The lattice constant determines the maximum number of layers in a bundle, $n=\lfloor l/a \rfloor$. If $l$ becomes too small for $n$ layers, the system loses the free energy gain from the Gaussian curvature of that $n^{\text{th}}$ layer. For experimental parameter values~\cite{hough09a,parameters} and $\k/B\sim(1{\rm{nm}})^2$, the ideal bundles have diameter $d\sim 20\rm{nm}$ and contain 5 layers.

\begin{figure}
\begin{center}
\includegraphics[width=80mm]{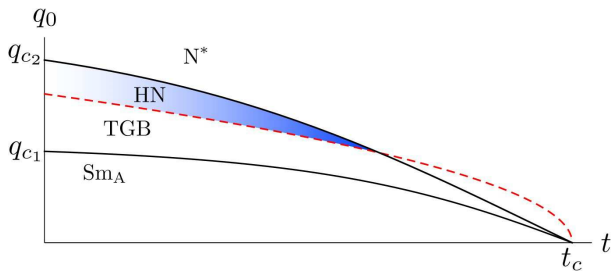}
\caption{(Color online) We show the schematic phase diagram for chiral smectics. The phase becomes unstable when the chirality exceeds $q_0=\alpha/\lambda \sim \vert t_c-t \vert^{1/2}$ (dashed, red line). The phase transition to the $\textrm{HN}$ phase occurs near the intersection between this line and the upper critical chirality $q_{c_2}$.}
\label{phasediagram}
\end{center}
\end{figure}

We can see directly from Eq.~\eqref{eq:hexenergy_numbers} that our bulk texture is energetically preferred over the smectic-A until a short distance below the thermodynamic critical chirality $q_{\text{th}}$, however its stability in relation to the $\TGB$ is less obvious. We focus on the region close to the upper critical chirality and argue that the $\TGB$ phase will lose stability relative to the HN phase for sufficiently large values of $q_0$. Within each grain of the TGB the compression energy locks the director to the local layer normal with a cosine potential, leading to a description for the twisting director field in terms of the infinite kink chain solution of the sine-Gordon model~\cite{manton04}. The width of the kinks (where the director rotates from one angle to the next) is set by the penetration depth $\lambda$. When the chirality is weak, $K_2 q_0^2  \ll B$, the separation between kinks $l_b$ is large and the rotation is confined to narrow regions, whereas when the chirality is large, $K_2 q_0^2 \gg B$, the rotation occurs at an almost uniform rate, reminiscent of the cholesteric director. As the transition is mean field continuous the $\TGB$ structure must evolve towards that of the high temperature $N^*$ phase as $q_{c_2}$ is approached. In particular, we have that $\tfrac{\alpha}{l_b} \nearrow q_0$ and that the director field crosses over from the step-wise rotation usually envisaged for the $\TGB$ phase to the uniform rotation of the cholesteric. 
Within the framework of the sine-Gordon model of the $\TGB$ director, this crossover is marked by the overlap of kinks when $l_b \lesssim \lambda$. In this regime the energy per unit volume of the $\TGB$ swiftly asymptotes to the value corresponding to having a cholesteric director field
\begin{equation}
\frac{F}{V} = - \frac{(t-t_c)^2}{u} + B \Bigl[ 1 - \frac{\sin (\alpha /2)}{\alpha /2} \Bigr] + B \frac{\sin (\alpha /2)}{q_{\text{sm}}l_b} \, \text{ln} \Bigl( \frac{l_d}{\xi} \Bigr) ,
\label{eq:TGB_energy}
\end{equation}
where $l_d$ is the spacing between screw dislocations within a grain boundary. As the chirality increases the constraint that $\tfrac{\alpha}{l_b} \nearrow q_0$ implies not only that the grain spacing $l_b$ decreases but also that the rotation angle $\alpha$ should increase and we can expect the high chirality limit to correspond to $\alpha \rightarrow \tfrac{\pi}{2}$. 
For such large grain angles $l_d = \sqrt{2}a$ and the dominant energy cost is provided by the second term of Eq.~\eqref{eq:TGB_energy}, taking a value $\sim 0.099 B$ that is larger than that of the HN phase. 
Thus for chiralites larger than $\sim \tfrac{\pi}{2\lambda}$ we expect the $\TGB$ phase to lose stability with respect to the HN phase, at least close to the upper critical chirality. The onset of the HN phase can be estimated from the intersection of the curves $q_0 = \tfrac{\pi}{2\lambda} \sim |t_c-t|^{1/2}$ and $q_0 = q_{c_2}(t) \sim |t_c-t|$ and hence should occur for $q_0 \gtrsim q_* = q_{\text{sm}} (\tfrac{\pi\xi}{2\lambda})^2$, leading to the schematic phase diagram of Fig.~\ref{phasediagram}.

We have proposed a new chiral organization of a smectic liquid crystal which fills space. Like the blue phases, this phase is stable at high chirality and is built out of smaller building blocks, in this case helical nanofilaments. We find that the compression energy has an instability towards ever finer structure. The fine structure is cut off by considering the discrete nature of the layers and their Gaussian curvature. This explains the fine scale of the bundles seen in experiment, where each bundle is only a few layers thick~\cite{hough09a}. Future work will examine the remarkable success of continuum theory despite the fact that the ``long'' lengthscales are comparable to the molecular size.  We will further explore the effects of order parameter variations and the deformations of the cholesteric texture.

\acknowledgements
We are grateful to Noel Clark for sharing experimental results with us prior to publication and thank B.G. Chen, D.J. Earl, G.M. Grason, R.D. James, and E. Tsai for beneficial discussions. This work was supported in part from NSF Grant DMR05-47320 and  gifts from L.J. Bernstein and H.H. Coburn. RDK acknowledges the hospitality of the Aspen Center for Physics where some of this work was completed.


\begin{thebibliography}{99}
\bibitem{filling} The corresponding result for non-overlapping bundles, $\epsilon < \tfrac{3}{4}$, is given by retaining only the first and third terms in each of the square brackets, but reduced by factors of $(4\epsilon/3)^2$ and $(4\epsilon/3)^7$, respectively.
\end{thebibliography}
\end{document}